# Improved Channel Estimation with Partial Sparse Constraint for AF Cooperative Communication Systems

Guan Gui and Wei Peng
Department of Communication Engineering
Tohoku University, Sendai, Japan
Email: {gui,peng}@mobile.ecei.tohoku.ac.jp

*Abstract*—Accurate channel state information (CSI) is necessary for coherent detection in amplify and forward (AF) broadband cooperative communication systems. Based on the assumption of ordinary sparse channel, efficient sparse channel estimation methods have been investigated in our previous works. However, when the cooperative channel exhibits partial sparse structure rather than ordinary sparsity, our previous method cannot take advantage of the prior information. In this paper, we propose an improved channel estimation method with partial sparse constraint on cooperative channel. At first, we formulate channel estimation as a compressive sensing problem and utilize sparse decomposition theory. Secondly, the cooperative channel is reconstructed by LASSO with partial sparse constraint. Finally, numerical simulations are carried out to confirm the superiority of proposed methods over ordinary sparse channel estimation methods.

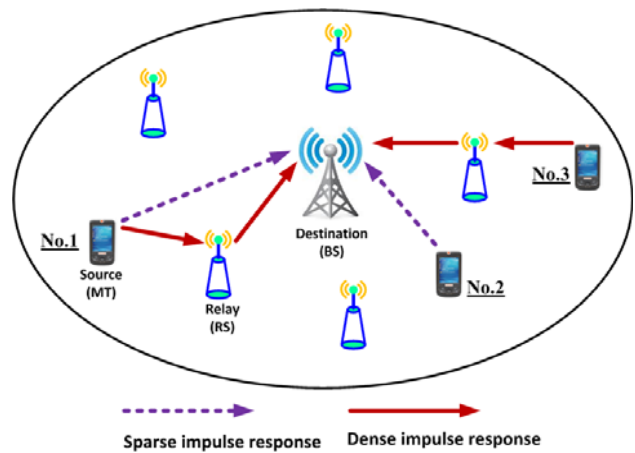

Fig.1. An example of AF broadband cooperative communication system, where source (MT) transmits signal to destination (BS) with the help of relay (RS). Since the RS can improve communication quality, the multipath channels are sparse and dense in direct link ( $\mathbb{S} \Rightarrow \mathbb{D}$ ) and cooperative link ( $\mathbb{S} \Rightarrow \mathbb{R} \Rightarrow \mathbb{D}$ ), respectively.

## I. INTRODUCTION

Relay-based cooperative communication systems (CCS) [1-4] have been studied in the last decade due to its capability of enhancing the transmission range and providing the spatial diversity for single-antenna receivers by employing the relay nodes as virtual antennas [5-7]. A typical example of cooperative communication system is shown in Fig.1. It is well known that utilizing multiple-inputs multiple outputs (MIMO) transmission can boost the channel capacity [8,9] in broadband communication systems. In addition, diversity techniques in MIMO system could mitigate selective fading and hence enhance the quality of service (QoS) [10,11]. However, it poses a practical challenge to integrate multiple antennas onto a small handhold terminal. To deal with the contradiction between them, one could choose relay-based cooperation networks which have been investigated in last decade [1,2,4]. The main reason is that the diversity from relay nodes existing in the network could be exploited, where relay can either be provided by operators or be obtained from cooperating terminals of other users.

In the relay-based cooperative communication system, data transmission is usually divided into two time slots. During first time slot, the source broadcasts its own information to both relay and destination. During second one, the relay could select different protocols and then transmit signal to the destination. Usually, there has two kinds of protocols in cooperative communication systems, one is amplify the received signal at relay and forward it to destination, which is termed as amplify-and-forward (AF); and the second is to decode the received signal, modulate it again, and then retransmit to destination, which is often termed as decode-and-forward (DF). Due to coherent detection in these systems, accurate channel state information (CSI) is required at the destination (for AF) or at both relay and destination (for DF). About DF cooperative communication systems, the channel estimation methods could be borrowed from point-to-point (P2P) communication systems directly. However, extra channel estimation will increase the computational burden at relay and broadcasting the estimated channel information will result in further interference at destination. On the other hand, AF cooperative communication technique can avoid this disadvantage and focus on AF CCS in this study.

Based on the theory of compressed sensing [12,13], sparse channel estimation methods [18-20] have been proposed for P2P communication systems. However, sparse channel estimation is also one of the key challenges in cooperative communication systems shown in Fig. 1. Linear channel estimation for the relay-based AF cooperative networks has been proposed [4] which is based on the assumption of dense multipath. Even though the proposed method can achieve lower bound performance, low spectrum efficiency is unavoidable since the utilized training sequence take a large space at the fixed bandwidth. As the channel measurement technique improves in the last decade, broadband wireless channels have been confirmed to exhibit inherent sparse or cluster-sparse structure in delay spread. One method to improve the spectrum efficiency is by reducing the number of training sequence for channel estimation. In order to take the advantages of channel's sparsity, we have proposed a sparse channel estimation scheme for ordinary sparse CCS [14] and global sparse constraint was considered in the proposed method. However, when the cooperative channel is partial sparse rather global sparse, the proposed method cannot fully take advantage of the prior information. In this paper, based on the partial sparse constraint, we propose a partial sparse channel estimation method by using LASSO [15] (PEL) to further exploit the channel prior information. Based on this idea, improved partial sparse channel estimation by using LASSO (IEL) is proposed by utilizing both partial sparse constraint and global sparse constraint. On one hand, partial sparse constraint can exploit partial sparse prior information. On the other hand, global sparse constraint can mitigate noise interference under low SNR. To confirm the effectiveness of the two proposed methods, we give various numerical simulation results in section IV.

Section II introduces the system model and problem formulation. In section III, two improved channel estimation methods are proposed. The first method is the improved channel estimation method by using partial sparse constraint and the second one is an improved partial sparse constraint. In section IV, we give various numerical simulation results and related discussions. Concluding remarks are presented in section V.

Notations: In this paper, we use boldface lower case letters $x$ to denote vectors, boldface capital letters $X$ to denote matrices. x represents the complex Gaussian random variable. E[.] stands for the expectation operation and $X$, $X^\dagger$ denote the matrix $X^H$ transposition and conjugated transposition operations. $\|x\|_0$ accounts the nonzero number of $x$ and $\|x\|_2$ is the Euclidean norm of $x$.

## II. SYSTEM MODEL

Consider a multipath fading AF broadband CCS where the source $\mathbb{S}$ sends data to destination $\mathbb{D}$ with the help of relay $\mathbb{R}$ as shown in Fig. 2. The three terminals are assumed to equip single antenna each. $h_{SD}$, $h_{SR}$ and $h_{RD}$ denote the impulse response of the frequency selective fading channel vectors between three links $\mathbb{S} \Rightarrow \mathbb{D}$, $\mathbb{S} \Rightarrow \mathbb{R}$ and $\mathbb{R} \Rightarrow \mathbb{D}$, respectively. Note that differ from our previous research in [14], impulse response of cooperative channels, $h_{SR}$ and $h_{RD}$ are modeled as dense channel model due to the fact that relay can reduce transmission range and improve channel quality. In other words, multipath taps arrive in a very short delay spread. The two channels are assumed to have length $L_{SR}$ and $L_{RD}$, respectively. For simplicity, we assume that they have same length $L_{SR} = L_{RD} = L/2$, and the channel model of $h_{SR}$ can be written as

$$h_{SR} = \sum_{l=0}^{L/2-1} h_{SR,l} \delta(t - \tau_{SR,l}), \quad (1)$$

where $h_{SR,l}$ and $\tau_{SR,l}$ represent the complex-valued path gain with $E[\sum_l |h_{SR,l}|^2] = 1$ and symbol spaced time delay of the $l$-th path, respectively. The training sequence vector $x$ is denoted as

$$x = [x(1), x(2), ..., x(N)]^T, \quad (2)$$

where the power constraint of the transmit power is $P_S = E[x^H x] = NP$, where $P$ is the unit transmitting power. According to the property of AF cooperative system which is shown in Fig.1, one full transmission can be divided into two time slots. At the first time slot, signal $x$ be broadcast the equivalent complex baseband received signal at $\mathbb{D}$ and $\mathbb{R}$ are given by

$$y_{D,1} = H_{SD} x + z_{D,1}, \quad (3)$$
$$y_{R,1} = H_{SR} x + z_{R,1}, \quad (4)$$

respectively, where $H_{SD}$ and $H_{SR}$ are $N \times N$ complex circulant channel matrices with its first columns $[h_{SD}^T, 0_{1 \times (N-L)}]^T$ and $[h_{SR}^T, 0_{1 \times (N-L/2)}]^T$ respectively; $z_{D,1}$ and $z_{R,1}$ is a realization of a complex additive Gaussian white noise vector with zero mean and covariance matrix $E[z_{D,1} z_{D,1}^H] = E[z_{R,1} z_{R,1}^H] = \sigma_n^2 I_N$. Then the relay $\mathbb{R}$ amplifies the received signal $y_{R,1}$ and retransmits the signal during the second time slot. The received signal vector at the destination $\mathbb{D}$ is given by

$$\begin{aligned} y_{D,2} &= \beta H_{RD} y_{R,1} + \tilde{z}_{D,2} \\ &= \beta H_{RD} H_{SR} x + z_{D,2}, \end{aligned} \quad (5)$$

where $H_{RD}$ is a circulant channel matrix with first column vector $[h_{RD}^T, 0_{1 \times (N-L/2)}]^T$; $z_{D,2} = \beta H_{RD} z_{R,1} + \tilde{z}_{D,2}$ is a composited AWGN with zero mean and covariance matrix $E[z_{D,2} z_{D,2}^H] = (\beta^2 |H_{RD}|^2 + I_N) \sigma_n^2$, where $\tilde{z}_{D,2}$ is a realization of a complex additive Gaussian white noise (AWGN) vector with zero mean and covariance matrix $\mathrm{E}[\tilde{z}_{D,2}\tilde{z}_{D,2}^H] = \sigma_n^2 I_N$. Considering long-time averaging, $\beta$ is given by

$$\beta = \sqrt{P_R / \sigma_h^2 P_S + \sigma_n^2}. \quad (6)$$

where $P_R$ is the transmit power of relay. Using Eq. (3) and Eq. (5), the effective input-output relation in the AF cooperative communication system can be summarized as

$$\tilde{y} = \begin{bmatrix} y_{D,1} \\ y_{D,2} \end{bmatrix} = \begin{bmatrix} H_{SD} \\ & \beta H_{SR} H_{RD} \end{bmatrix} \begin{bmatrix} x \\ x \end{bmatrix} + \begin{bmatrix} z_{D,1} \\ z_{D,2} \end{bmatrix}. \quad (7)$$

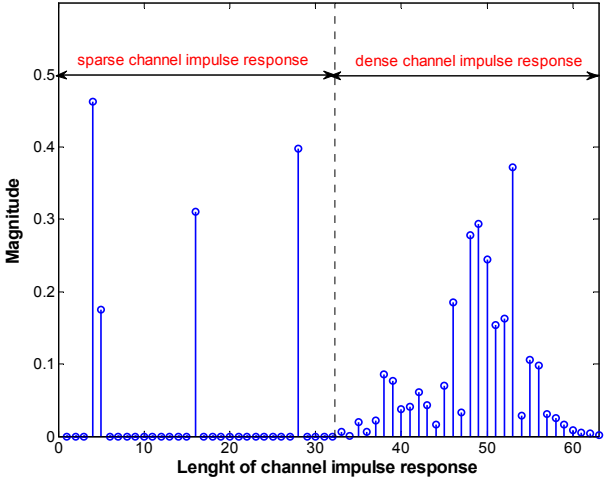

Fig.2. A typical example of partial sparse cooperative channel, where the first part of sparse impulse response is supported by direct link and the second part of dense impulse response is contributed by cooperative cascaded link.

According to matrix theory [22], all circulant matrices can share the same eigenvectors. That is to say, the same unitary matrix can work for all circulant matrices. Hence, the matrices $H_{SD}$, $H_{SR}$ and $H_{RD}$ in Eq. (7) are de-composed as $H_{SD} = F^H D_{SD} F$, $H_{SR} = F^H D_{SR} F$, and $H_{RD} = F^H D_{RD} F$, respectively, where $F$ is the unitary discrete Fourier transform (DFT) matrix with entries $f_{mn} = [F]_{mn} = 1/\sqrt{N} e^{-j2\pi(m-1)(n-1)/N}$, $m,n = 1,2,...,N$. Hence, cooperative channel matrix $H_{SR} H_{RD}$ can be written as

$$H_{SR} H_{RD} = F^H D_{SRD} F, \quad (8)$$

where $D_{SRD} = D_{SR} D_{RD}$ denotes a diagonal matrix. At the same time, $F^H D_{SRD} F$ is the decomposition of a circulant matrix which is constructed from a cascaded channel impulse response $h_{SRD} \triangleq h_{SR} * h_{RD}$. $D_{SD}$ and $D_{SRD}$ are diagonal matrices which are given by

$$D_{SD} = diag\{H_{SD}(0),...,H_{SD}(n),...,H_{SD}(N-1)\}, \quad (9)$$

$$D_{SRD} = diag\{H_{SRD}(0),...,H_{SRD}(n),...,H_{SRD}(N-1)\}, \quad (10)$$

respectively, where $H_{SD}(n)$ and $H_{SRD}(n)$ are given by

$$H_{SD}(n) = \sum_{l=0}^{L-1} h_{SD}(l) e^{-j2\pi nl/N}, \quad (11)$$

$$H_{SRD}(n) = \sum_{l=0}^{L-2} h_{SRD}(l) e^{-j2\pi nl/N}, \quad (12)$$

respectively, where $h_{SD} = [h_{SD}(0), h_{SD}(1),...,h_{SD}(L-1)]^T$ denotes direct link from source $\mathbb{S}$ to destination $\mathbb{D}$ at the first time slot and $h_{SRD} = [h_{SRD}(0), h_{SRD}(1),...,h_{SRD}(L-2)]^T$ represents cascaded channel from source $\mathbb{S}$ to destination $\mathbb{D}$ via help of the relay $\mathbb{R}$ at the second time slot. Based on the above analysis, Eq. (7) is left multiplied by $F$ and it can be rewritten as

$$y = Xh + z, \quad (13)$$

where $y = [(Fy_{D,1})^T, (Fy_{D,2})^T]^T$ denotes $2N$-length received signal vector; $X$ denotes equivalent training matrix and it can be written as

$$X = \begin{bmatrix} Fdiag(x)F_{SD} & 0_{N\times(L-1)} \\ 0_{N\times L} & Fdiag(x)F_{SRD} \end{bmatrix}, \quad (14)$$

with $2N \times (2L-1)$ dimension; $h = [h_{SD}^T \; h_{SRD}^T]^T$ represents $(2L-1)$-length cooperative channel vector; $z = [(Fz_{D,1})^T \; (Fz_{D,2})^T]^T$ denote $2N$-length complex AWGN vector; $F_{SD}$ and $F_{SRD}$ are partial DFT matrices taking the first $L$ and $(L-1)$ columns of $F$, respectively. And the $z$ is a realization of a complex Gaussian random vector with zero mean and covariance matrix $\mathrm{E}[zz^H] = (\beta^2 |D_{SR}|^2 + I_{2N})\sigma_n^2$.

## III. HIGH-RESOLUTION COMPRESSIVE CHANNEL ESTIAMTION

In this section, we discuss partial sparse channel estimation for AF CCS. At first, we review briefly CS theory and restricted Isometry property (RIP) of training signal matrix. Then, we propose improved sparse channel estimators by using partial sparse constraint.

### A. Review of the CS

In a typical complex sparse identification system, one can use known matrix $U \in \mathbb{C}^{N \times L}$ to estimate a $L$-length unknown sparse signal vector $a$ based on the observation linear system model

$$b = Ua + c, \quad (15)$$

where $b \in \mathbb{C}^N$ is a complex observation signal vector, $c \in \mathbb{C}^N$ is a noise vector, and $a$ is $K$ sparse vector which means the number of dominant entries is no more than $K$, i.e., $\|a\|_0 \leq K \ll L$. The position of dominant entries is randomly distributed. In addition, $L \gg N$ according to CS assumption.

Mathematically, the optimal sparse solution $a_{opt}$ can be obtained uniquely by solving minimization problem

$$a_{opt} = \arg\left\{\min_a \|a\|_0, \; subject \; to \; \|b - Ua\|_2^2 \leq \xi\right\}, \quad (16)$$

where $\xi \geq 0$ denotes noise error tolerance. Above minimization problem in Eq. (16) is also equivalent to

$$a_{opt} = \arg\lim_a \left\{\frac{1}{2}\|b - Ua\|_2^2 + \lambda_0 \|a\|_0\right\}, \quad (17)$$

where $\lambda_0$ is regularized parameter which tradeoffs the mean square error (MSE) and sparsity. However, solving $\ell_0$ norm is NP hard and cannot be utilized in practical applications [12].

Fortunately, alternative sub-optimal sparse recovery methods have been studied if the known measurement matrix $U$ satisfies RIP [23]. Let $U_\Omega$, $\Omega \subset \{1,2,...,N\}$ be the $N \times |\Omega|$ submatrix extracting those columns of $U$ that are indexed by the elements of $\Omega$. Then the $K$-restricted Isometry constant (RIC) of $U$ is defined as the smallest parameter $\delta_K \in (0,1)$ such that

$$\left|\frac{\|U_\Omega a_\Omega\|_2^2 - \|a_\Omega\|_2^2}{\|a_\Omega\|_2^2}\right| \leq \delta_K, \quad (18)$$

for all $\Omega$ with $|\Omega| \leq K$ and all vector $a_\Omega \in \mathbb{C}^{|\Omega|}$.

*Theorem 1* [16]: Assume that $U$ is an $N \times L$ random measurement matrix that satisfies the RIP of order $K$ with RIC $\delta_K$, that is $U \in \mathrm{RIP}(K, \delta_K)$. Consider an arbitrary sparse vector $a$ in observation model $b = Ua + c$, where $\|c\|_2 \leq \xi$, by solving $\ell_1$ minimization problem and sub-optimal sparse solution $\hat{a}_{sub}$ is obtained by

$$\hat{a}_{sub} = \arg\min_a \left\{\frac{1}{2}\|b - Ua\|_2^2 + \lambda_{sub} \|a\|_1\right\}, \quad (19)$$

where $\lambda_{sub} = C_0 \cdot \sigma_n \log N$ and $C_0$ is a parameter which is decided by the noise level and RIC of $U$. Hence, the estimator $\hat{a}_{sub}$ satisfies sparse recovery performance with

$$\|\hat{\boldsymbol{a}}_{sub} - \boldsymbol{a}\|_2 \le C_1 \max\left\{\xi, 1/\sqrt{K}\ \|\boldsymbol{a} - \boldsymbol{a}_K\|_1\right\} \quad (20)$$

where $C_1$ is a parameter which is also decided by noise level and RIC of $\boldsymbol{U}$. Let us recall the channel estimation problem for AF cooperative systems in Eq. (13), if the equivalent training matrix $\boldsymbol{X}$ satisfies RIP, then accurate sparse channel estimation can be achieved. In the next, we will present improved sparse channel estimation methods by using LASSO algorithm [15].

*B. Partial sparse channel estimation*

In this section, channel estimation is done on partial sparse channel $\boldsymbol{h}$ by sending the training symbols. Conventional sparse channel estimation method using LASSO algorithm (SEL) has been proposed for deriving sparse impulse response for AF CCS [14]. According to the system model in Eq. (13), the ordinary sparse channel estimator $\hat{\boldsymbol{h}}_{SEL}$ can be achieved by

$$\hat{\boldsymbol{h}}_{SEL} = \arg\min_{\boldsymbol{h}} \left\{ \frac{1}{2}\|\boldsymbol{y} - \boldsymbol{X}\boldsymbol{h}\|_2^2 + \lambda_{SEL} \|\boldsymbol{W}_{SEL}\boldsymbol{h}\|_1 \right\}, \quad (21)$$

where

$$\boldsymbol{W}_{SEL} = \boldsymbol{I}_{(2L-1)\times(2L-1)}, \quad (22)$$

is a identity matrix and $\lambda_{SEL} > 0$ is a regularization parameter which decides the tradeoff between mean square error $\|\boldsymbol{y} - \boldsymbol{X}\boldsymbol{h}\|_2^2$ and sparse constraint $\|\boldsymbol{W}_{SEL}\boldsymbol{h}\|_1$.

The above estimation method can solve global sparse solution well while neglecting the inherent partial sparse structure. From signal processing perspective, extra prior information can be further utilized. In this situation, partial sparse channel estimation by using LASSO (PEL) $\hat{\boldsymbol{h}}_{PEL}$ could be achieved by

$$\hat{\boldsymbol{h}}_{PEL} = \arg\min_{\boldsymbol{h}} \left\{ \frac{1}{2}\|\boldsymbol{y} - \boldsymbol{X}\boldsymbol{h}\|_2^2 + \lambda_{PEL} \|\boldsymbol{W}_{PEL}\boldsymbol{h}\|_1 \right\}, \quad (23)$$

where

$$\boldsymbol{W}_{PEL} = \begin{bmatrix} \boldsymbol{I}_{L\times L} & \boldsymbol{0}_{L\times(L-1)} \\ \boldsymbol{0}_{(L-1)\times L} & \boldsymbol{0}_{(L-1)\times(L-1)} \end{bmatrix}, \quad (24)$$

is a diagonal weighted matrix and $\lambda_{PEL} > 0$ is a regularization parameter which controls the trade-off between square error $\|\boldsymbol{y} - \boldsymbol{X}\boldsymbol{h}\|_2^2$ and partial sparse constrained $\|\boldsymbol{W}_{PEL}\boldsymbol{h}\|_1$.

Based on the partial sparse constraint on cooperative channel impulse response, we consider an improved PEL (IEL) estimator. On one hand, local sparse constrain on cooperative can improve estimation performance. On the other hand, global sparse constraint can mitigate noise interference in the low SNR regime. The IEL estimator $\hat{\boldsymbol{h}}_{IEL}$ can be obtained by

$$\hat{\boldsymbol{h}}_{IEL} = \arg\min_{\boldsymbol{h}} \left\{ \frac{1}{2}\|\boldsymbol{y} - \boldsymbol{X}\boldsymbol{h}\|_2^2 + \lambda_{SEL} \|\boldsymbol{W}_{SEL}\boldsymbol{h}\|_1 + \lambda_{PEL} \|\boldsymbol{W}_{PEL}\boldsymbol{h}\|_1 \right\}, \quad (25)$$

where the regularization parameters $\lambda_{SEL}$ and $\lambda_{PEL}$ are given by the Eq. (21) and Eq. (24), respectively. In the following, we will give various simulation results to confirm the effectiveness of the improved sparse channel estimation methods.

## IV. NUMERICAL SIMULATIONS

In this section, we will compare the performance of the proposed PEL estimator and IEL estimator with traditional methods: SEL estimator and LS estimator. To achieve average estimation performance, 10000 independent Monte-Carlo runs are adopted. The length of training sequence is $N = 36$. The length of direct link $\boldsymbol{h}_{SD}$ is $L = 32$ with number of dominant channel taps $K = 2, 4, 8$ and cooperative channels $\boldsymbol{h}_{SR}$ and $\boldsymbol{h}_{RD}$ are $L/2$-length dense impulse response. All of the nonzero channel taps are generated following Rayleigh distribution and subject to

$$\mathrm{E}[\|\boldsymbol{h}_{SR}\|_2] = \mathrm{E}[\|\boldsymbol{h}_{RD}\|_2] = \mathrm{E}[\|\boldsymbol{h}_{SD}\|_2] = 1. \quad (26)$$

Transmit power and AF relay power are fixed as $P_S = P_R = NP$, where $P$ is a unit transmit power. The received SNR is defined as $P_S / \sigma_n^2$. Channel estimator $\hat{\boldsymbol{h}}$ is evaluated by average mean square error (average MSE) which is defined by

$$Average\ MSE(\hat{\boldsymbol{h}}) = \frac{\mathrm{E}\left[\|\boldsymbol{h} - \hat{\boldsymbol{h}}\|_2^2\right]}{2L-1}, \quad (27)$$

where $\boldsymbol{h}$ and $\hat{\boldsymbol{h}}$ denote composite channel vector and its estimator, respectively; $(2L-1)$ is the overall length of channel vector $\boldsymbol{h}$. At first, we compare their estimation performance with different number of dominant channel taps in the direct link of cooperative channel. When the number of dominant channel taps in direct link is $K = 2$, two proposed channel estimators (PEL and IEL) are better than SEL and LS based estimator as shown in Fig. 3. From the figure, we can find that IEL estimator has a better performance than PEL under low SNR (less than 15dB). However, when the number of dominant channel taps increase, the advantage of IEL reduces and its performance is close to PEL as shown in Figs. 4 and 5.

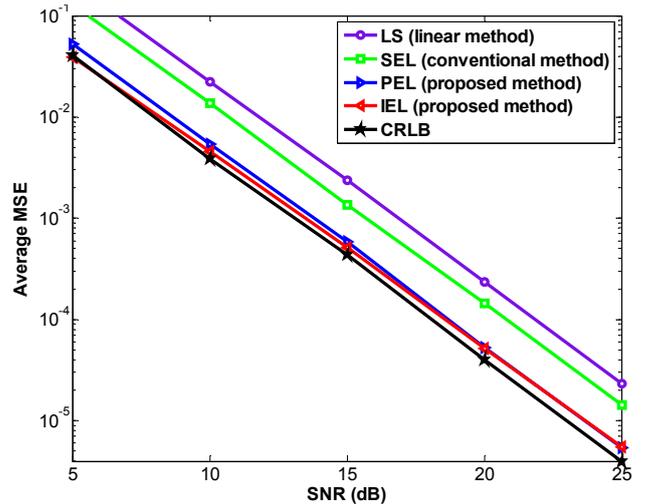

Fig.3. Channel estimation performance versus SNR. The number of dominant channel taps in direct link is $K = 2$.

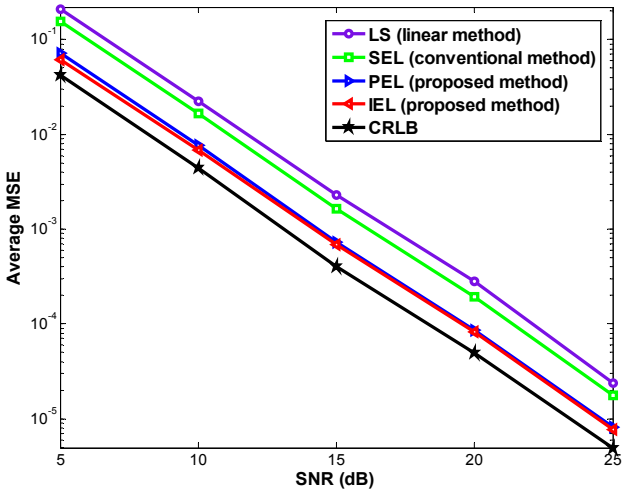

Fig.4. Channel estimation performance versus SNR. The number of dominant channel taps in direct link is $K = 4$.

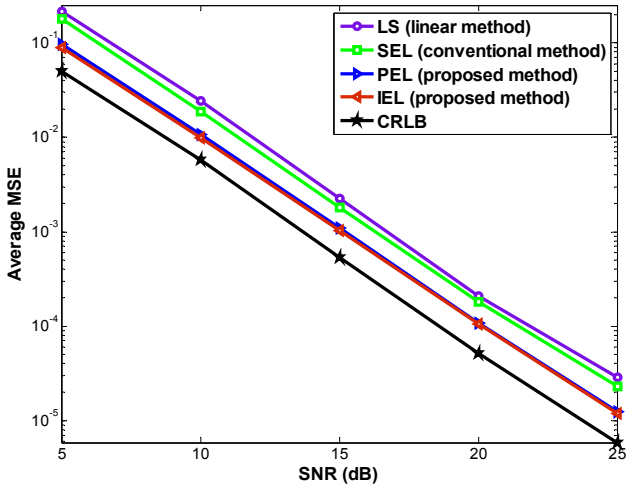

Fig.5. Channel estimation performance versus SNR. The number of dominant channel taps in direct link is $K = 8$.

On sparse channel estimation, channel sparsity often affects the channel estimation performance. That is to say, sparser channel can achieve more accurate estimation performance under the same condition. We evaluate the performance of the channel estimators (PEL and IEL) with the different number of dominant channel taps $K = 2, 4, 8$ as shown in Figs. 6 and 7. When the number of dominant channel taps is very small, e.g., $K = 2$, the estimation performance is close to the lower bound. However, if the direct link is dense channel impulse response, then the two proposed estimator are close to LS-based channel estimator. At the same time, if the cascaded link is sparse channel impulse response then the two proposed estimator have same performance as SEL. According to the above analysis, we can find that the proposed methods are generalized from both LS-based and SEL, since they are either based on dense or sparse channel assumption. Hence, our proposed methods in this paper can work well in different channel environments.

In the next, the relationship between the proposed method and channel sparsity is considered. Since the space limitation, here, we only consider the uniform distribution of dominant taps on compressive channel estimation. Assume that the number of channel length is same, while the number of dominant taps is 2, 4, 6 and 8, respectively. We also compare their recovery probability of dominant channel taps (see Fig. 9) and average MSE (see Fig. 10). From the two figures, whatever the channel sparsity, our proposed method can always close to their lower bound. Hence, the proposed method is stable for the sparse channel with different number of dominant channel taps.

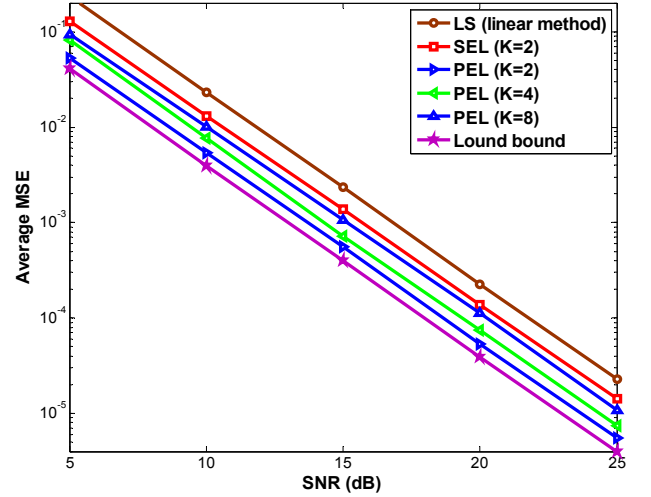

Fig.6. PEL channel estimators' performance versus the number of dominant channel taps in direct link of cooperative channel.

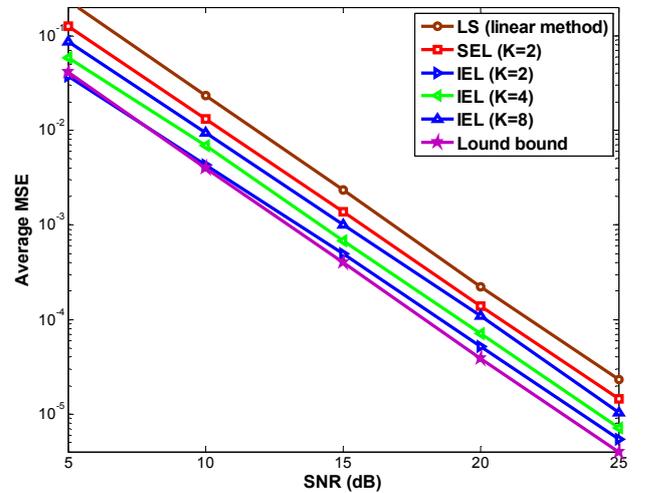

Fig.7. IEL channel estimators' performance versus the number of dominant channel taps in direct link of cooperative channel.

## V. CONCLUSION

Traditional channel estimation methods are based on assumptions of either dense channel model or sparse channel model in AF CCS. In this paper, the two kinds of channel models have been generalized as a partial sparse channel. By means of compressive sensing and partial sparse constraint, we have proposed an improved sparse channel estimation method to fully exploit channel prior information. Numerical simulations have confirmed the performance superiority of the proposed method to the conventional ordinary sparse channel estimation method and traditional linear LS method.